# Statistical theory of self–similar time series

A. I. Olemskoi

*Department of Physical Electronics, Sumy State University, Rimskii-Korsakov St. 2, 40007 Sumy, Ukraine*[*]

(Dated: October 30, 2002)

Within Tsallis statistics, a picture is elaborated to address self–similar time series as a thermodynamic system. Thermodynamic–type characteristics relevant to temperature, pressure, entropy, internal and free energies are introduced and tested. Predictability conditions of time series analysis are discussed in details on the basis of Van der Waals model. Maximal magnitude for time interval and minimal resolution scale of the value under consideration are found and analyzed in details. Time series statistics is shown to be governed by effective temperature being exponential measure of the fractal dimensionality of a phase space related to the time series.

## I. INTRODUCTION

Time series analysis allows to elaborate and verify macroscopic models of complex systems evolution on the basis of data analysis [1]. This analysis is known to be focused on numerical calculations of correlation sum for delay vectors that allow to find principle characteristics of the time series [2]. Being traditionally a branch of the theory of statistics, time series analysis is based on the class of models of harmonic oscillator which are related to the simplest case of the Gaussian random process [3]. But well known real time series is relevant rather to the Lévy stable processes, than the Gaussian ones being very special case [4]. Because the formers are invariant with respect to dilatation transformation [5], the problem is reduced to consideration of self–similar stochastic processes.

The simplest characteristic of time series is known to be the Lyapunov exponent which maximum positive value yields predictability domain in the system behavior. A range of complexity in such behavior is determined by the Kolmogorov–Sinai entropy that equals to sum over positive magnitudes of whole set of the Lyapunov exponents and may be reduced to the usual Shannon value in information theory. With passage to nonlinear system the probability $p_i$ of $i$–th scenario of the system behavior is transformed into power function $p_i^q$ determined by index $q \leq 1$, so that the Kolmogorov–Sinai entropy is replaced by the Renyi one

$$K_q \equiv \frac{\ln \sum_i p_i^q}{1 - q}. \qquad (1)$$

Respectively, governing equation $\mathrm{d}p/\mathrm{d}\epsilon = -\beta p^q$, $\beta = \mathrm{const} > 0$ describes probability variation with energy $\epsilon = \epsilon_i$ to arrive at the Tsallis distribution [6]

$$p \propto [1 - (1-q)\beta\epsilon]^{\frac{1}{1-q}}. \qquad (2)$$

In the limit $q \to 1$, this distribution takes usual Boltzmannian form $p \propto \exp(-\beta\epsilon)$ falling down exponentially fast in contrary to the power asymptotic of the generalized exponent (2). Physically, such a behavior is caused by self–similarity of the system related to the Tsallis statistics [7] — [10]. As a result, the problem appears to study self–similar time series that present processes corresponded to power–law distribution type of Zipf–Mandelbrot ones [11]. This work is devoted to analytical consideration of such type time series as a thermodynamic system.

The paper is organized as follows. Section II is devoted to elaboration of a model which permits to address a self–similar time series as a nonextensive thermodynamic system. Section III is based on calculations of both entropy and internal energy of the time series. As a result, thermodynamic–type characteristics of the time series such as temperature and entropy, volume and pressure, internal and free energies are introduced. Their testing for the model of ideal gas is shown to be basis for statistics of self–similar time series. Section IV contains calculations of corrections to the ideal gas approach when external field and particle interaction are switched on. On the basis of Van der Waals model, we obtain the expressions for the specific heat and susceptibility as functions of the temperature. Section V is devoted to discussion of the physical meaning of the results obtained within the framework where the predictability of behavior of time series is mimicked as stability conditions of a non–ideal gas. We find maximal magnitudes for time interval and minimal resolution scale of the value under consideration. Concluding Section VI shows that a temperature governing time series statistics is exponential measure of a self–similarity index related to fractal dimensionality of the phase space. Finally, several equalities needed in quoting are placed in Appendix.

## II. STATEMENT OF TIME SERIES AS NONEXTENSIVE THERMODYNAMIC SYSTEM

Let us consider $d$–dimensional time series $\mathbf{x}(t_n)$ related to the set $\{\mathbf{x}_n\}$ of consequent values $\mathbf{x}_n \equiv \mathbf{x}(t_n)$ of principle variable $\mathbf{x}(t)$ taken at discrete time instants $t_n \equiv n\tau$ that we obtain as result of dividing a whole time series length $\mathcal{T} \equiv N\tau$ by $N$ equal intervals $\tau$. It is obviously to be relevant to the time series $\mathbf{x}(t_n)$, the set $\{\mathbf{x}_n\}$ should be supplemented by conjugated set $\{\mathbf{v}_n\}$ of velocities, which show rates of $\mathbf{x}_n$–variation with the time jumping. In the simplest case of Markovian consequence, one has $\mathbf{v}_n \equiv (\mathbf{x}_n - \mathbf{x}_{n-1})/\tau$. For more complicated series with $m$-step memory, the velocity magnitude is defined as follows:

$$v_n \equiv \sqrt{\frac{1}{m}\sum_{i=1}^{m} \vec{\delta}_i^{\,2}(m,n)}, \qquad (3)$$

$$\vec{\delta}_i(m,n) \equiv \frac{\mathbf{x}_{(n-m)+i} - \mathbf{x}_{(n-m)+(i-1)}}{\tau}.$$

---

[*]Electronic address: olemskoi@ssu.sumy.ua

The paradigm of our approach is to address the time series as a physical system defined by an effective Hamiltonian $\mathcal{H} = \mathcal{H}\{\mathbf{x}_n, \mathbf{v}_n\}$ on whose basis statistical characteristics of this series could be found. If one proposes that series terms $\mathbf{x}_n$ related to different $n$ are not connected, the effective Hamiltonian is additive:

$$\mathcal{H} = \sum_{n=1}^{N} \varepsilon_n, \quad \varepsilon_n \equiv \varepsilon(\mathbf{x}_n, \mathbf{v}_n). \qquad (4)$$

Physically, this means that the series under consideration is relevant to an ideal gas comprising of $N$ identical particles with energies $\varepsilon_n$. Further, we suppose different terms of time series to be statistically identical, so that effective particle energy does not depend on coordinate $\mathbf{x}_n$: $\varepsilon(\mathbf{x}_n, \mathbf{v}_n) \Rightarrow \varepsilon(\mathbf{v}_n)$. Moreover, since this energy does not vary with inversion of the coordinate jumps $\mathbf{x}_n - \mathbf{x}_{n-1}$, the function $\varepsilon(\mathbf{v}_n)$ should be even. We use the simplest square form

$$\varepsilon_n = \frac{1}{2}\mathbf{v}_n^2, \qquad (5)$$

which is reduced to the usual kinetic energy for a particle with mass 1. With switching on an external force $\mathbf{F} = \mathbf{const}$, particle energy (5) becomes as follows:

$$\varepsilon_n = \frac{1}{2}\mathbf{v}_n^2 - \mathbf{F}\mathbf{x}_n. \qquad (6)$$

Finally, when time series has a microscopic memory, dimension of the delay vectors $\vec{\delta}_i(m,n)$ in the definition (3) needs taking $m > 1$. Moreover, if time series terms $\mathbf{x}_m$, $\mathbf{x}_n$ with $m \neq n$ are clustered, the Hamiltonian becomes relevant to a non–ideal gas with interaction $w_{mn}$, $m \neq n$:

$$\mathcal{H} = \frac{1}{2}\sum_{n=1}^{N}\mathbf{v}_n^2 + \frac{1}{2}\sum_{m\neq n} w_{mn}. \qquad (7)$$

To study a behavior of the time series as a whole one needs to fulfil summation over a set of states given by manifold $\{\mathbf{x}_n, \mathbf{v}_n\}$ that is relevant to the system phase space. In so doing, it is convenient to pass to related integrations as following:

$$\sum_{\{\mathbf{x}_n,\mathbf{v}_n\}} \Rightarrow \iint \prod_{n=1}^{N} \frac{\mathrm{d}\mathbf{x}_n \mathrm{d}\mathbf{v}_n}{N!\Delta} = \mathcal{N}^{-1}\prod_{n=1}^{N}\iint \mathrm{d}\mathbf{y}_n \mathrm{d}\mathbf{u}_n. \qquad (8)$$

Here, the factorial takes into account statistical identity of the time series terms, $\Delta$ is effective Planck constant that determines a minimal volume of the phase space per a particle related to a term. The inverted factor

$$\mathcal{N} \equiv N!\left(\frac{X^2}{\tau\Delta}\right)^{-dN} \simeq \left[\frac{\mathrm{e}X^{2d}}{N(\tau\Delta)^d}\right]^{-N} \qquad (9)$$

is caused by change of variables

$$\mathbf{y}_n \equiv \frac{\mathbf{x}_n}{X}, \quad \mathbf{u}_n \equiv \frac{\tau\mathbf{v}_n}{X} \qquad (10)$$

rescaled with respect to macroscopic length $X$ being chosen to guarantee the conditions

$$\int \mathrm{d}\mathbf{y}_n = 1, \quad \int \mathrm{d}\mathbf{u}_n = 1. \qquad (11)$$

According to Introduction self–similarity condition forces to use a statistics type of given by the Tsallis distribution (2). However, the latter is appeared to be inconvenient because it is normalized with nonconventional condition

$$\sum_i p_i^q \equiv \langle 1 \rangle_q \neq 1 \qquad (12)$$

and does not take into account the definition of the internal energy

$$\sum_i \frac{\epsilon_i p_i^q}{\langle 1 \rangle_q} = E. \qquad (13)$$

To avoid these difficulties an escort distribution was introduced [12]

$$\mathcal{P}_i \equiv \frac{p_i^q}{\langle 1 \rangle_q}. \qquad (14)$$

In explicit form it reads as follows:

$$\mathcal{P}_q\{\mathbf{y}_n, \mathbf{u}_n\} = \begin{cases} \frac{1}{Z}\left[1 - (1-q)\frac{\mathcal{H}\{\mathbf{y}_n,\mathbf{u}_n\}-E}{\langle 1 \rangle_q T_s}\right]^{\frac{q}{1-q}} & \text{at} \quad (1-q)\frac{\mathcal{H}\{\mathbf{y}_n,\mathbf{u}_n\}-E}{\langle 1 \rangle_q T_s} < 1, \\ 0 & \text{otherwise.} \end{cases} \qquad (15)$$

Here, the partition function is defined by condition

$$Z \equiv \mathcal{N}^{-1} \prod_{n=1}^{N} \iint \left[1 - (1-q)\frac{\mathcal{H}\{\mathbf{y}_n, \mathbf{u}_n\} - E}{\langle 1 \rangle_q T_s}\right]^{\frac{q}{1-q}} \mathrm{d}\mathbf{y}_n \mathrm{d}\mathbf{u}_n, \quad (16)$$

where $0 < q < 1$ is a parameter of nonextensivity, $T_s$ is energy scale. Internal energy $E$ is determined by equality

$$E \equiv \mathcal{N}^{-1} \prod_{n=1}^{N} \iint \mathcal{H}\{\mathbf{y}_n, \mathbf{u}_n\} \mathcal{P}_q\{\mathbf{y}_n, \mathbf{u}_n\} \mathrm{d}\mathbf{y}_n \mathrm{d}\mathbf{u}_n, \quad (17)$$

and normalization parameter $\langle 1 \rangle_q$ is expressed by the partition function (16) in accordance with Eq.(79).

To check the statistical scheme proposed let us address firstly trivial case of time series $\mathbf{x}_n = \mathbf{const}$. Here, the particle energy $\varepsilon$ is a constant as well, so that the Hamiltonian is $\mathcal{H} = N\varepsilon$. The partition function $Z = \mathcal{N}^{-1}$ and the normalization parameter $\langle 1 \rangle_q = \mathcal{N}^{-(1-q)}$ are given by inverted normalization factor (9), whereas the internal energy $E = N\varepsilon$ is reduced to the Hamiltonian. Then, the entropy $H = -\ln\mathcal{N}$ obtained according to definition (77) given in Appendix is reduced to zero if only the normalization factor takes the value $\mathcal{N} = 1$. As a result, we find effective Planck constant:

$$\Delta = \left(\frac{\mathrm{e}}{N}\right)^{\frac{1}{d}} \frac{X^2}{\tau}. \quad (18)$$

Our future consideration is stated on the assumption that the volume $V \equiv X^d$ of $d$-dimensional domain of the coordinate $\mathbf{x}_n$ variation in dependence of the particle number $N$ is governed by Lévy–type law

$$X^d = x^d N^{\frac{1}{z}}. \quad (19)$$

Here, $x$ is a microscopic constant and a dynamic exponent $z$ is reduced to the fractal dimensionality $D$ of self–similar phase space [8]

$$z = D. \quad (20)$$

Then, one obtains the following scaling relation for the phase space volume per a term of the time series:

$$\Delta^d = \mathrm{e}\left(\frac{x^2}{\tau}\right)^d N^{\frac{2}{D}-1}. \quad (21)$$

In the case of Gaussian scattering, when $D = 2$, the minimal volume $\Delta^d$ of the phase space does not depend on number $N$ of the time series terms. Such a condition approves our choice of the relation (19) for the whole volume $V \equiv X^d$ as function of the number $N$ of time series terms.

### III. NONEXTENSIVE THERMODYNAMICS OF TIME SERIES AS AN IDEAL GAS

Calculations of main thermodynamic quantities of nonextensive ideal gas arrives at the following expressions for the partition function (16) and the internal energy (17) (see [12] — [14])

$$Z = \frac{V^N \gamma(q)}{N!} \left[\frac{\theta \langle 1 \rangle_q}{1-q}\right]^{\frac{dN}{2}} \left[1 + (1-q)\frac{dN}{2}\right]^{-1} \left[1 + (1-q)\frac{E}{\langle 1 \rangle_q T_s}\right]^{\frac{1}{1-q}+\frac{dN}{2}}, \quad (22)$$

$$E = \frac{dN}{2} \frac{V^N \gamma(q) T_s}{N!} \left[\frac{\theta \langle 1 \rangle_q}{1-q}\right]^{\frac{dN}{2}} \left[1 + (1-q)\frac{dN}{2}\right]^{-1} \left[1 + (1-q)\frac{E}{\langle 1 \rangle_q T_s}\right]^{\frac{1}{1-q}+\frac{dN}{2}} Z^{-q}. \quad (23)$$

Here, $d$-dimensional gas in volume $V \equiv X^d$ is addressed and the notations are introduced

$$\theta \equiv \frac{2\pi T_s}{\Delta^2}, \quad \gamma(q) \equiv \frac{\Gamma\left(\frac{1}{1-q}\right)}{\Gamma\left(\frac{1}{1-q} + \frac{dN}{2}\right)} \quad (24)$$

to be determined by the Euler $\Gamma$–function. Combination of equalities (22), (23) with relation (79) yields explicit expression for the normalization parameter:

$$\langle 1 \rangle_q = \left\{\frac{X^{dN} \gamma(q)}{N!} \left[\frac{\theta(1+a)^{\frac{q}{a}+1}}{1-q}\right]^{\frac{dN}{2}}\right\}^{\frac{1-q}{1-a}} \quad (25)$$

$$a \equiv \frac{1}{2}(1-q)dN. \tag{26}$$

In the limits

$$1 - q \ll 2/d, \quad N \gg 1 \tag{27}$$

when

$$\gamma(q) \simeq [\mathrm{e}(1-q)]^{\frac{dN}{2}} (1+a)^{-\frac{1+a}{1-q}}, \tag{28}$$

one obtains for the entropy (77):

$$H \simeq \frac{Na}{2(1-a)} \ln\left[\mathrm{e}^{2+d}\theta^d \left(\frac{X^d}{N}\right)^2\right]. \tag{29}$$

With accounting scaling relation (19), this expression takes the usual form

$$H = N\frac{D-1}{D} \ln\left(\frac{G}{N}\right), \quad G \equiv (2\pi\mathrm{e}T_s)^{\frac{dD}{2}} \left(\frac{x}{\tau}\right)^{-dD} \tag{30}$$

if the dynamic exponent is determined as

$$z \equiv D = \frac{1}{1-a}. \tag{31}$$

Respectively, the internal energy (23) and the normalization parameter (25) read:

$$E = \frac{dN}{2}\left(\frac{G}{N}\right)^{\frac{2a}{d}} T_s, \ \langle 1 \rangle_q = \left(\frac{G}{N}\right)^{\frac{2a}{d}}; \ a \equiv \frac{D-1}{D}. \tag{32}$$

The physical temperature is defined as follows [13]

$$T \equiv \langle 1 \rangle_q T_s = \left(\frac{G}{N}\right)^{\frac{2a}{d}} T_s \tag{33}$$

where the last equality takes into account the second of relations (32). This definition guarantees the equipartition law

$$E = CT, \quad C \equiv cN, \quad c \equiv \frac{d}{2} \tag{34}$$

where the quantity

$$C = \frac{\partial E}{\partial T} \tag{35}$$

is the specific heat. It is easily to convince that equations (30) — (33) arrive at standard thermodynamic relation

$$\frac{\partial H}{\partial E} \equiv \frac{1}{T}. \tag{36}$$

Above used treatment is addressed to a fixed value of the internal energy $E$ [14]. In alternative case when the principle state parameter is the temperature $T$, we should pass to the conjugate formalism [12]. Here, standard definition

$$F \equiv E - TH \tag{37}$$

of the free energy arrives at the dependence

$$F = -CT \ln\left(\frac{T}{\mathrm{e}T_s}\right). \tag{38}$$

Then, the thermodynamic identity

$$\frac{\partial F}{\partial T} \equiv -H \tag{39}$$

yields the relation

$$H = C \ln\left(\frac{T}{T_s}\right) \tag{40}$$

that plays a role of the heat equation of states. It arrives at the usual definition of the specific heat (cf. Eq. (35))

$$C = T\frac{\partial H}{\partial T}. \tag{41}$$

Let us introduce now specific entropy per unit time

$$h \equiv (d\tau)^{-1}\frac{\partial H}{\partial N} = \tau^{-1}H_1 - r \tag{42}$$

to be determined by a minimal entropy

$$H_1 = (D-1)\left[\ln\sqrt{2\pi\mathrm{e}T_s} - \ln\left(\frac{x}{\tau}\right)\right] \tag{43}$$

and a redundancy

$$r = \frac{a}{d\tau}\ln(\mathrm{e}N), \quad a \equiv \frac{D-1}{D}. \tag{44}$$

Dependencies on the scale $x$

$$h(x) = \mathrm{const} - \frac{D-1}{\tau}\ln\left(\frac{x}{\tau}\right), \quad r(x) = \mathrm{const} \tag{45}$$

notice that the system behaves in a stochastic manner [15].

Effective pressure is defined as

$$p \equiv -\tau\frac{\partial h}{\partial x} \tag{46}$$

to measure specific entropy variation with respect to the time series scale. Then, we arrive at a mechanic–type equation of states

$$px = D - 1 \tag{47}$$

being additional to the relation (40) of entropy to temperature. According to Eq. (47), definition of the pressure coefficient

$$\kappa \equiv \frac{\partial p}{\partial (x^{-1})} \tag{48}$$

shows that it is fixed by the dynamic exponent (31) being the fractal dimensionality of self–similar phase space:

$$\kappa = D - 1. \tag{49}$$

It is worthwhile to note that effective pressure (46) is introduced as derivative of the specific entropy $h$ with respect to the microscopic scale $x$. This means a microscopic nature of so defined pressure which arrives at a susceptibility of the time series with respect to a choice of the microscopic scale $x$ inverted (obviously, this susceptibility is inversion of the pressure coefficient (48)).

## IV. CORRECTIONS TO THE IDEAL GAS APPROACH

Let us focus now on effect of external field and particle interaction which switching on is expressed by equalities (6), (7). An external force $\mathbf{F} = \mathbf{const}$ causes the second term in Hamiltonian (6) to arrive at the factor

$$Z_{ext} = Z_d \left[ \frac{\sinh\left(\frac{FX}{2T}\right)}{\frac{FX}{2T}} \right]^N \qquad (50)$$

in partition function (22). Here, $Z_d$ is a factor depended on the dimensionality $d$ only and we put $q \to 1$ due to the conditions (27). According to the definition (77), the factor (50) yields the entropy addition

$$H_{ext} = Na \ln \left[ \frac{\sinh\left(\frac{FX}{2T}\right)}{\frac{FX}{2T}} \right] \qquad (51)$$

where we suppress unessential term. With increasing homogeneous external field, the entropy $H_{ext}$ grows quadratically at $FX \ll T$ and linearly at $FX \gg T$.

According to [16], cluster expansion of the particle interaction in Hamiltonian (7) results in additional factor in partition function (22):

$$Z_{int} = 1 - \frac{N^2}{2V}\left(v + \frac{w}{T}\right); \quad w \equiv S_d \int_\epsilon^\infty w(x) x^{d-1} dx, \; S_d \equiv \frac{2\pi^{d/2}}{\Gamma(d/2)}; \quad v \equiv \frac{S_d}{d}\epsilon^d \qquad (52)$$

where $\epsilon$ is effective radius of the particle core. Relevant entropy addition

$$H_{int} = -\frac{aN^2}{2V}\left(v + \frac{w}{T}\right) \qquad (53)$$

monotonously decreases with growth of the particle volume $v$. Increasing temperature $T$ causes the entropy decrease for attractive interaction $w < 0$ and its increase in the case of repelling one ($w > 0$).

Entropy additions (51), (53) arrive at total value of the specific heat (41) in the following form:

$$C = \frac{d}{2}N - aN\left\{ \left[\frac{FX}{2T}\coth\left(\frac{FX}{2T}\right) - 1\right] - \frac{N}{2V}\frac{w}{T}\right\} \qquad (54)$$

where formulae (34) take into account. On the other hand, making use of equalities (42), (46), (51), (53) arrives at the pressure addition $\delta p$ determined by the equality

$$\delta p \cdot x = -\frac{a}{d}\left[\frac{FX}{2T}\coth\left(\frac{FX}{2T}\right) - 1\right] + a\frac{N}{V}\left(v + \frac{w}{T}\right). \qquad (55)$$

Then, with accounting (49) the pressure coefficient (48) takes the form

$$\kappa = \frac{a}{1-a} + \frac{a}{d}\left\{1 + \left[\frac{\frac{FX}{2T}}{\cosh\left(\frac{FX}{2T}\right)}\right]^2\right\} + (1+d)an\left(v + \frac{w}{T}\right) \qquad (56)$$

where effective density

$$n \equiv \frac{N}{V} = x^{-d}N^a \qquad (57)$$

is introduced. In the limiting case of low temperatures $T \ll FX/2$, we obtain:

$$C \simeq \frac{d}{2}N - \frac{aN}{2T}(FX - nw), \qquad (58)$$

$$\kappa \simeq \frac{a}{1-a} + \frac{a}{d}\left[1 + \left(\frac{FX}{T}\right)^2 \exp\left(-\frac{FX}{T}\right)\right] + (1+d)an\left(v + \frac{w}{T}\right). \qquad (59)$$

Respectively, in opposite case $T \gg FX/2$, one has:

$$C \simeq \frac{d}{2}N + \frac{aN}{2T}\left[nw - \frac{(FX)^2}{6T}\right], \qquad (60)$$

$$\kappa \simeq \frac{a}{1-a} + \frac{a}{d}\left[1 + \frac{1}{4}\left(\frac{FX}{T}\right)^2\right] + (1+d)an\left(v + \frac{w}{T}\right). \qquad (61)$$

Thus, in the limits of both low and high temperatures, the influence of external field $F$ reduces to hyperbolically decreasing addition $(FX/T)^n$, $n = 1, 2$ to the specific heat $C$; accordingly, the pressure coefficient $\kappa$ gets really the constant $a/d$. In a similar manner, the particle interaction affects hyperbolically, as $w/T$ with intensity determined by interaction parameter $w$, on the specific heat, whereas term being proportional to the particle volume $v$ is added to the pressure coefficient.

## V. STABILITY CONDITIONS

It might seem the analysis of a time series would appeared to be the subject of more much study if both the interval $\tau$ and the scale $x$ take infinitely decreasing magnitudes (respectively, the number $N$ tends to infinitely large values). However, we will show further that external influence and term clustering arrive at predictabilities boundaries in behavior of relevant system that are represented as instability of modelling thermodynamic system. Our analysis is stated on stability conditions

$$C > 0, \quad \kappa > 0 \qquad (62)$$

for the specific heat $C$ and the pressure coefficient $\kappa$ given by equalities (54) and (56).

In the simplest approach being modelled by the ideal gas in Section III, we arrives at the natural restriction for the dynamic exponent:

$$z \equiv D > 1. \qquad (63)$$

It means nonextensive dependence of the system volume (19) on the particle number $N$. On the other hand, in accordance with relation (31) the magnitude of the principle index (26) is limited by condition

$$a < 1 \qquad (64)$$

which restricts the rate of increasing effective density (57) with $N$.

As is seen from expressions (54), (58), (60), the specific heat falls down monotonically with temperature decrease. To analyze such a behavior quantitatively let us consider the case of low temperatures $T \ll FX/2$, when we can use the simplest of above dependencies (58).

Then, it is easily to see that at temperature less than critical magnitude

$$T_c = \frac{a}{d}\left(FxN^{\frac{1-a}{d}} - wx^{-d}N^a\right) \qquad (65)$$

the system becomes nonstable due to external force $F$ and interparticle attraction $w < 0$. If the number $N$ is much less than critical value $N_c$ defined as

$$N_c^{\frac{1}{d} - \frac{1+d}{d}a} \equiv \frac{|w|}{F}x^{-(1+d)}, \qquad (66)$$

main contribution gets interparticle attraction $w < 0$, in opposite case $N \gg N_c$ — external force $F$. With passage to more complicated case of high temperatures $T \gg FX/2$, when one follows to use temperature dependence (60), the physical situation does not change qualitatively.

As show estimations (59), (61), the pressure coefficient becomes negative if microscopic scale $x^d$ is less than a critical magnitude

$$x_c^d = \frac{d(1+d)}{1 + \frac{d}{1-a}}\left(\frac{|w|}{T} - v\right)N^a \qquad (67)$$

that is determined at attractive nature of interaction only ($w < 0$). If both parameters $w$ and $v$ takes nonzeroth magnitudes, the temperature values are characterized by another minimal magnitude $T_v = |w|/v$, lower which the microscopic scale $x_c$ becomes principle. This temperature is less than critical value (65) if elementary volume takes magnitudes lower than value

$$v_c \equiv \frac{d}{a}n^{-1} = \frac{d}{a}x^d N^{-a} \qquad (68)$$

where we take into account definition (57).

In the limits $v \to 0$, $F \to 0$, boundary magnitudes (65), (67) arrive at appearance of the critical value

$$a_c = \frac{(1+d)(1+d^2)}{2}\left[1 - \sqrt{1 - \frac{4d^2}{(1+d)(1+d^2)^2}}\right]. \qquad (69)$$

Relevant fractal dimensionality of the phase space (31) grows monotonically with dimensionality of the time series $d$ from the magnitude $D_c = 1 + \sqrt{2} \simeq 2.414$ at $d = 1$ to $D_c \simeq d^2$ at $d \gg 1$ (see Figure 1). As a result, the critical value (69) arrives at the upper boundary of the fractal dimensionality $D_c$ being more than topologic magnitude $D_t = 2d$ always ($D_c > D_t$). Thus, we can conclude the upper boundary $D_c$ of application of our approach is appeared to be inessential at all.

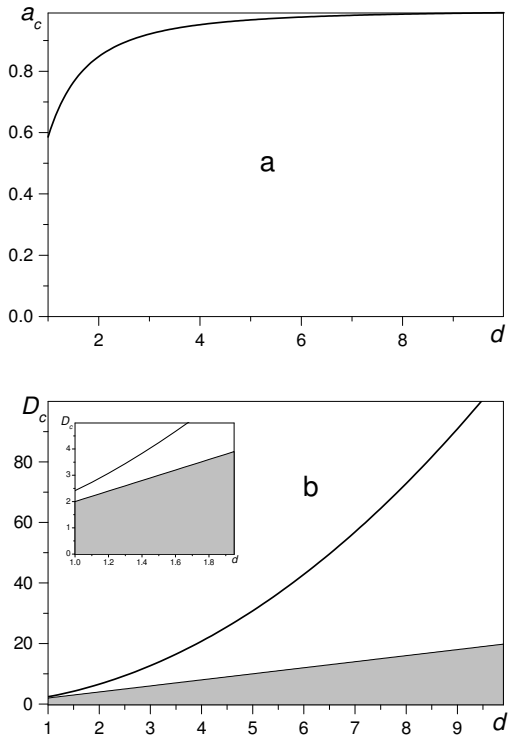

FIG. 1: a. The dependence of the critical value (69) on the time series dimensionality. b. The same for related maximal value (31) of the fractal dimensionality of the phase space (solid line) and topologic magnitude $D_t = 2d$ (thin line). The physical domain is shaded.

## VI. DISCUSSION

We have addressed above the simplest model which have allowed us to examine analytically a self–similar time series in standard statistical manner. Characteristic peculiarity of related equalities is a scale invariance with respect to variation of the nonextensivity parameter $1 - q$ which is contained everywhere as combination (26) derived to the parameter $a$ that is related to the dynamic exponent $z$ and the fractal fractal dimensionality $D$ of the phase space according to Eq. (31). This invariance is clear to be caused by self–similarity of the system under consideration. For a given time series, the value $D$ is fixed if it is addressed to a monofractal manifold and takes a closed set of magnitudes in the case of self–similar system relevant to a multifractal.

In real time series the property of self–similarity can be broken, so that a dependence on the nonextensivity parameter itself could appear to be very weak. However, above introduced set of pseudo–thermodynamic characteristics of time series is kept as applicable and accustomed thermodynamic relations (34) — (37), (39), (41), (46) and (48) can be applied to analysis of arbitrary time series.

Main progress in our consideration is that time series statistics is governed completely by the temperature (33). With accounting Eq.(30) and rescaling the temperature unit $T_s$ into $T_{sc} \equiv (2\pi\mathrm{e})^a T_s$ we derive to the expression

$$\frac{T}{T_{sc}} = \left[\left(\frac{\tau}{X}\right)^2 T_{sc}\right]^{\frac{a}{1-a}}. \tag{70}$$

Being independent of the number of terms $N$, the time series temperature shows exponential dependence on the index (26) located under critical magnitude (69). To establish character of the power dependence on the ratio of the range $X$ of the principle variable to the time interval $\tau$, it is naturally to choice measure units of the temperature in the following manner:

$$T_{sc} \equiv \mathrm{e}\left(\frac{X}{\tau}\right)^2, \quad T_s \equiv \frac{\mathrm{e}^{1-a}}{(2\pi)^a}\left(\frac{X}{\tau}\right)^2 \approx \frac{1}{2\pi}\left(\frac{X}{\tau}\right)^2. \tag{71}$$

Then, the expression (70) for the time series temperature takes the simplest form

$$T = \left(\frac{X}{\tau}\right)^2 \mathrm{e}^D, \quad D \equiv \frac{1}{1-a} \tag{72}$$

according to which the value $T$ is exponential measure of the fractal dimensionality $D$ of the self–similar phase space related.

As has shown consideration in Section V, a time series subject external influence and term clustering is limited in predictability of behavior of relevant system that is appeared as instability of modelling thermodynamic system. On the basis of Van der Waals model, we have found minimal magnitudes (67), (68) for resolution scale $x_c$ of the value under consideration. Respectively, making use of the critical temperature (65) and the definition (72) shows that a time series is predictable if the microscopic time interval $\tau$ takes magnitudes less than a critical magnitude $\tau_c$. In the case $N \ll N_c$, where the boundary magnitude is determined by Eq. (66), we find

$$\tau_c = \left(\frac{a}{d}|w|\right)^{-\frac{1}{2}} \mathrm{e}^{\frac{D}{2}} x^{1+\frac{d}{2}} N^{\frac{2+d}{2Dd}-\frac{1}{2}}. \tag{73}$$

In opposite case $N \gg N_c$, one obtains

$$\tau_c = \left(\frac{a}{d}F\right)^{-\frac{1}{2}} \mathrm{e}^{\frac{D}{2}} x^{\frac{1}{2}} N^{\frac{1}{2Dd}}. \tag{74}$$

It is easily to convince that the entropy (40) takes positive values within the stability domain $\tau < \tau_c$.

To conclude, we point out that one of advantages of the approach proposed is a possibility of its application to numerical analysis of real time series. The basis of this calculations is stated on expressions (3) — (7), (15) — (17), (33), (34), (35), (37), (39), (41), (42), (46), (48) and (77). This work is in progress.

**Acknowledgments**

Author is grateful to Max–Planck–Institute for Physics of Complex Systems (Dresden, Germany) for warm hospitality which allows to start this work.

**Appendix: Main statements of the nonextensive statistics**

Our approach is stated on using entropy definitions

$$H = a\ln[\exp_q(H_q)], \quad H_q = \ln_q[\exp(H/a)]; \quad (75)$$
$$a \equiv \tfrac{1}{2}(1-q)dN$$

that are alternations of the usual functions logarithm $\ln(x)$ and exponential $\exp(x)$ with corresponding Tsallis generalizations [6]

$$\ln_q(x) \equiv \frac{x^{1-q}-1}{1-q}, \quad \exp_q(x) \equiv [1+(1-q)x]^{\frac{1}{1-q}}. \quad (76)$$

In explicit form, relations (75) are appeared as

$$H \equiv a\ln Z, \quad H_q \equiv \frac{\langle 1\rangle_q - 1}{1-q}. \quad (77)$$

Making use of the first equality (75) and the second formulae (76) and (77) shows that physically defined entropy $H$ is extensive value being reduced to the Renyi definition (1):

$$H = aK_q = \frac{Nd}{2}\ln\langle 1\rangle_q. \quad (78)$$

Comparison of this result with the first of equalities (77) arrives at the relation

$$\langle 1\rangle_q = Z^{1-q} \quad (79)$$

that ensures the normalization condition [6].

[1] G. Boffetta, M. Cencini, M. Falcioni, A. Vulpiani, *Phys. Rep.* **356**, 367 (2002).
[2] H. Kantz, T. Schriber, *Nonlinear time series analysis* (Cambridge University Press, Cambridge, 1997).
[3] P.J. Brockwell, R.A. Davis, *Springer Texts in Statistics. Introduction to Time Series and Forecasting* (Springer, New York, 1998).
[4] R.N. Mantegna, H.E. Stanley, *An Introduction to Econophysics. Correlations and Complexity in Finance* (Cambridge University Press, Cambridge, 2000).
[5] D. Sornette, *Critical Phenomena in Natural Sciences* (Springer, Berlin, Heidelberg, 2000).
[6] C. Tsallis, in: *Lecture Notes in Physics*, S. Abe, Y. Okamoto, Eds. (Springer-Verlag, Heidelberg, 2001).
[7] D. H. Zanette, P. A. Alemany, *Phys. Rev. Lett.* **75**, 366 (1995).
[8] D. H. Zanette, *Braz. J. Phys.* **29**, 108 (1999); cond–mat/9905064.
[9] A. I. Olemskoi, *JETP Let.* **71**, 285 (2000).
[10] A. I. Olemskoi, *Physica A* **295**, 409 (2001).
[11] B.B. Mandelbrot, *The Fractal Geometry of Nature* (Freeman, New York, 1983).
[12] C. Tsallis, R.S. Mendes, A.R. Plastino, *Physica A* **261**, 534 (1998).
[13] S. Abe, S. Martínez, F. Pennini, A. Plastino, *Phys. Lett. A* **281**, 126 (2001).
[14] S. Martínez, F. Nicolás, F. Pennini, A. Plastino, *Physica A* **286**, 489 (2000).
[15] M. Cencini, M. Falcioni, E. Olbrich, H. Kantz, A. Vulpiani, *Phys. Rev. E* **62**, 427 (2000).
[16] S. Martínez, F. Pennini, A. Plastino, *Phys. Lett. A* **282**, 263 (2001).